\documentclass[12pt]{article}
\usepackage{amssymb}

\input{tcilatex}

\begin{document}

\bigskip\ 

\bigskip\ 

\begin{center}
\textbf{THE }$2+2$\textbf{-SIGNATURE AND THE }$1+1$\textbf{-MATRIX-BRANE}

\smallskip\ 

\smallskip\ 

J. A. Nieto\footnote{%
nieto@uas.uasnet.mx}

\smallskip\ 

\textit{Facultad de Ciencias F\'{\i}sico-Matem\'{a}ticas de la Universidad
Aut\'{o}noma}

\textit{de Sinaloa, 80010, Culiac\'{a}n Sinaloa, M\'{e}xico}

\bigskip\ 

\bigskip\ 

\textbf{Abstract}
\end{center}

We discuss different aspects of the $2+2$-signature from the point of view
of the quatl theory. In particular, we compare two alternative approaches to
such a spacetime signature, namely the $1+1$-matrix-brane and the $2+2$%
-target spacetime of a string. This analysis also reveals hidden discrete
symmetries of the $2+2$-brane action associated with the $2+2$-dimensional
sector of a $2+10$-dimensional target background.

\bigskip\ 

\bigskip\ 

\bigskip\ 

\bigskip\ 

\bigskip\ 

\bigskip\ 

\bigskip\ 


Pacs numbers: 04.60.-m, 04.65.+e, 11.15.-q, 11.30.Ly

November 29, 2006

\newpage

\noindent \textbf{1.- Introduction}

\smallskip\ 

Through the years it has become evident that the $2+2$-signature is not only
mathematically interesting [1]-[2] (see also Refs. [3]-[5]) but also
physically. In fact, the $2+2-$signature emerges in several physical
context, including self-dual gravity \textit{a la} Plebanski (see Ref. [6]
and references therein), consistent $N=2$ superstring theory as discussed by
Ooguri and Vafa [7], $N=(2,1)$ heterotic string [8]-[12]. Moreover, it has
been emphasized [13]-[14] that Majorana-Weyl spinor exists in spacetime of $%
2+2-$signature.

More recently, using the requirement of the $SL(2,R)$ and Lorentz symmetries
it has been proved [15] that $2+2$-target spacetime of a $0$-brane is an
exceptional signature. Moreover, following an alternative idea to the notion
of worldsheets for worldsheets proposed by Green [16] or the $0-$branes
condensation suggested by Townsend [17] it was also proved in Ref. [15] that
special kind of $0$-brane called quatl [18]-[19] leads to the result that
the $2+2$-target spacetime can be understood either as $2+2$-worldvolume
spacetime or as $1+1$-matrix-brane.

Another recent motivation for a physical interest in the $2+2$-signature has
emerged via the Duff's [20] discovery of hidden symmetries of the Nambu-Goto
action. In fact, this author was able to rewrite the Nambu-Goto action in a $%
2+2$-target spacetime in terms of a hyperdeterminant, reveling apparently
new hidden symmetries of such an action.

Our main goal in this brief note is to establish a connection between the
formalisms of references [15] and [20]. Specifically, we contrast the
similarities and the differences between these two works. In particular, we
construct the matrix linking the $2+2$-target spacetime of the two scenarios.

\bigskip\ 

\noindent \textbf{2.- The signatures} $2+2$ \textbf{in the quatl theory}

\smallskip\ 

In this section we shall briefly review the quatl theory focusing in the $%
2+2 $-signature. For that purpose consider the line element

\begin{equation}
ds^{2}=d\xi ^{A}d\xi ^{B}\eta _{AB}.  \tag{1}
\end{equation}%
Here, we shall assume that the indices $A,B\in \{1,2,3,4\}$ and that $\eta
_{AB}=diag(1,1,-1,-1)$ determines the $2+2$-signature$.$ By defining

\begin{equation}
\begin{array}{ccccccc}
\zeta ^{11}\equiv \xi ^{1}, &  & \zeta ^{22}\equiv \xi ^{2}, &  & \zeta
^{12}\equiv \xi ^{3}, &  & \zeta ^{21}\equiv \xi ^{4},%
\end{array}
\tag{2}
\end{equation}%
it is not difficult to show that (1) can also be written as

\begin{equation}
ds^{2}=d\zeta ^{am}d\zeta ^{bn}\eta _{ab}\eta _{mn},  \tag{3}
\end{equation}%
where $a,b,m,n\in \{1,2\},$ and $\eta _{ab}=diag(1,-1)$ and $\eta
_{mn}=diag(1,-1)$. We see that while $\xi ^{A}$ are coordinates associated
with a `spacetime' of signature $2+2$ the coordinates $\zeta ^{am}$ are
associated with a `spacetime' of signature $1+1.$ Thus, the equivalence
between the line elements (1) and (3) determines an interesting connection
between the signatures $1+1$ and $2+2$.

The $2+2$-brane action is given by

\begin{equation}
S=\frac{1}{2}\tint d\xi ^{2+2}\sqrt{-G}\left[ G^{AB}\frac{\partial x^{\hat{%
\nu}}}{\partial \xi ^{A}}\frac{\partial x^{\hat{\sigma}}}{\partial \xi ^{B}}%
\gamma _{\hat{\nu}\hat{\sigma}}-2\right] ,  \tag{4}
\end{equation}%
where $G_{AB}=G_{BA}$ is an auxiliary metric and $\gamma _{\hat{\nu}\hat{%
\sigma}}=\gamma _{\hat{\sigma}\hat{\nu}}$ is a metric in a higher
dimensional target spacetime.

The action (4) leads to the constraint

\begin{equation}
\frac{\partial x^{\hat{\nu}}}{\partial \xi ^{A}}\frac{\partial x^{\hat{\sigma%
}}}{\partial \xi ^{B}}\gamma _{\hat{\nu}\hat{\sigma}}-\frac{G_{AB}}{2}\left[
G^{CD}\frac{\partial x^{\hat{\nu}}}{\partial \xi ^{C}}\frac{\partial x^{\hat{%
\sigma}}}{\partial \xi ^{D}}\gamma _{\hat{\nu}\hat{\sigma}}-2\right] =0, 
\tag{5}
\end{equation}%
from which we can derive the expression

\begin{equation}
G^{CD}\frac{\partial x^{\hat{\nu}}}{\partial \xi ^{C}}\frac{\partial x^{\hat{%
\sigma}}}{\partial \xi ^{D}}\gamma _{\hat{\nu}\hat{\sigma}}=4.  \tag{6}
\end{equation}%
Using (6) one sees that (5) yields

\begin{equation}
\frac{\partial x^{\hat{\nu}}}{\partial \xi ^{A}}\frac{\partial x^{\hat{\sigma%
}}}{\partial \xi ^{B}}\gamma _{\hat{\nu}\hat{\sigma}}=G_{AB},  \tag{7}
\end{equation}%
and therefore the action (4) is reduced to the Nambu-Goto action.

It turns out that the constraint (5), and consequently the action (4), can
be obtained as a first quantization of the quatl action [15] (see also Refs.
[18] and [19])

\begin{equation}
\mathcal{S}=\frac{1}{2}\tint d\tau \left\{ \dot{\xi}^{A}p_{A}-\sqrt{-G}\left[
G^{AB}p_{A}p_{B}-\frac{2}{S+T}\right] \right\} .  \tag{8}
\end{equation}%
Here, $S$ and $T$ denotes the number of time and space coordinates of the
target spacetime associated with the quatl, respectively. Further, $p_{A}$
is the canonical momentum associated with the coordinate $\xi ^{A}$.

Now, both actions (4) and (8) can be written in terms of coordinates $\zeta
^{am}$ as

\begin{equation}
S=\frac{1}{2}\tint d\zeta ^{1+1}\sqrt{-g}\sqrt{-\gamma }\left[ g^{ab}\gamma
^{mn}\frac{\partial x^{\hat{\nu}}}{\partial \zeta ^{am}}\frac{\partial x^{%
\hat{\sigma}}}{\partial \zeta ^{an}}\gamma _{\hat{\nu}\hat{\sigma}}-2\right]
,  \tag{9}
\end{equation}%
and

\begin{equation}
\mathcal{S}=\tint d\tau \left[ \dot{\zeta}^{am}p_{am}-\frac{1}{2}\sqrt{-g}%
\sqrt{-\gamma }(g^{ab}\gamma ^{mn}p_{am}p_{bn}-\frac{2}{S+T})\right] , 
\tag{10}
\end{equation}%
respectively. Here, $g_{ab}(\zeta ^{am})$ and $\gamma _{mn}(\zeta ^{am})$
are two different nonsymmetric metrics (see Ref. [15] for details) and $%
p_{am}$ is the canonical momentum associated with the coordinate $\zeta ^{am}
$. The system described by the action (9) is called `$1+1-$matrix-brane'
system.

\bigskip\ 

\noindent \textbf{3.- Nambu-Goto action in a }$2+2$-\textbf{dimensional
target spacetime}

\smallskip\ 

Let us assume a string moving in a $2+2$-dimensional\textbf{\ }target
spacetime determined by the flat metric $\eta _{\mu \nu }=dig(1,1,-1,-1).$
The Nambu-Goto action for this system is given by

\begin{equation}
S=\tint d\xi ^{1+1}\sqrt{-h},  \tag{11}
\end{equation}%
where $h$ is the determinant of the matrix

\begin{equation}
h_{ab}=\frac{\partial x^{\mu }}{\partial \xi ^{a}}\frac{\partial x^{\nu }}{%
\partial \xi ^{b}}\eta _{\mu \nu }.  \tag{12}
\end{equation}

Duff [20] discover that by introducing the matrix

\begin{equation}
x^{pq}=\left( 
\begin{array}{cc}
x^{1}+x^{3} & x^{4}+x^{2} \\ 
x^{4}-x^{2} & x^{1}-x^{3}%
\end{array}%
\right) ,  \tag{13}
\end{equation}%
the action (11) can be written as

\begin{equation}
S=\tint d\xi ^{1+1}\sqrt{-Det(h_{ab})},  \tag{14}
\end{equation}%
where

\begin{equation}
Det(h_{ab})\equiv \frac{1}{2!}\varepsilon ^{ic}\varepsilon ^{jd}\varepsilon
^{ef}\varepsilon ^{gh}\varepsilon ^{rs}\varepsilon
^{lm}v_{egi}v_{fhj}v_{rlc}v_{smd},  \tag{15}
\end{equation}%
with

\begin{equation}
v_{pqa}=\frac{\partial x_{pq}}{\partial \xi ^{a}}.  \tag{16}
\end{equation}%
One of the advantages of writing the Nambu-Goto action as in (14) rather
than as (11) is that a number of hidden discrete symmetries can be revealed
(see Ref. [20] for details).

\bigskip\ 

\noindent \textbf{4.- A relation between the formalisms of sections 2 and 3}

\smallskip\ 

In this section we shall discuss a number of ways how we can link the
approaches of the previous sections 2 and 3. For this purpose it turns out
convenient to define three different kinds of `spacetimes', namely the
target spacetime ($\mathcal{T}$), the scenario where a $0-$brane moves, the
worldvolume ($\mathcal{W}_{s+t}$) associated with a $s+t-$brane, and the $%
S+T $ background target ($\mathcal{BT}_{S+T}$) spacetime where the $t+s-$%
brane evolves.

In section 2, we considered a connection of the form

\begin{equation}
\mathcal{T\leftrightsquigarrow W}_{2+2}\mathcal{\leftrightsquigarrow BT}%
_{S+T},  \tag{17}
\end{equation}%
which can be obtained after quantizing the quatl system. Here, the symbol $%
\mathcal{\leftrightsquigarrow }$ means a connection. We also mentioned that: 
$\mathcal{W}_{2+2}$ can be related to the $1+1$-signature. Let us express
this result in the form

\begin{equation}
\mathcal{W}_{2+2}=\mathcal{W}_{(1+1)+(1+1)}.  \tag{18}
\end{equation}

On the other hand the development in section 3 can be summarized
symbolically as

\begin{equation}
\mathcal{W}_{1+1}\mathcal{\leftrightsquigarrow BT}_{2+2}\mathcal{%
\leftrightsquigarrow BT}_{(1+1)+(1+1)}.  \tag{19}
\end{equation}%
It is worth mentioning that (19) is not rigorously a correct connection
since quantum consistency of the string theory establishes that the correct
link is

\begin{equation}
\mathcal{W}_{1+1}\mathcal{\leftrightsquigarrow BT}_{S+T=26}.  \tag{20}
\end{equation}

With the help of the symbolic connections (17)-(19) it comes to be evident
that a link between the formalisms of sections 2 and 3 can be established by
the simultaneous projections

\begin{equation}
\mathcal{W}_{(1+1)+(1+1)}\rightarrow \mathcal{W}_{1+1},  \tag{21}
\end{equation}%
and

\begin{equation}
\mathcal{BT}_{(1+2)+(1+2)}\rightarrow \mathcal{BT}_{(1+1)+(1+1)},  \tag{22}
\end{equation}%
which can be achieved through the so called double dimensional reduction
[21]. Therefore, in principle we have the scenario

\begin{equation}
\mathcal{T\leftrightsquigarrow W}_{2+2}\mathcal{\leftrightsquigarrow W}%
_{(1+1)+(1+1)}\rightarrow \mathcal{W}_{1+1}\mathcal{\leftrightsquigarrow BT}%
_{(1+2)+(1+2)}\rightarrow \mathcal{BT}_{(1+1)+(1+1)}.  \tag{23}
\end{equation}%
This means that form a first quantization of the quatl system one can obtain
the theory explained in section 3.and expressed symbolically in (19).

\bigskip\ 

\noindent \textbf{4.- Alternative relation between the formalisms of
sections 2 and 3}

\smallskip\ 

In this section we shall use the Duff prescription of section 3 to describe
an alternative but equivalent formalism for the quatl theory in $2+2$
dimensions. For this purpose let us write the line element (1) in the
alternative form

\begin{equation}
ds^{2}=\frac{1}{2}d\lambda ^{am}d\lambda ^{bn}\varepsilon _{ab}\varepsilon
_{mn},  \tag{24}
\end{equation}%
where

\begin{equation}
\lambda ^{pq}=\left( 
\begin{array}{cc}
\xi ^{1}+\xi ^{3} & \xi ^{4}+\xi ^{2} \\ 
\xi ^{4}-\xi ^{2} & \xi ^{1}-\xi ^{3}%
\end{array}%
\right) .  \tag{25}
\end{equation}

We have

\begin{equation}
\frac{\partial x^{\hat{\nu}}}{\partial \xi ^{A}}=\frac{\partial x^{\hat{\nu}}%
}{\partial \lambda ^{pq}}\frac{\partial \lambda ^{pq}}{\partial \xi ^{A}}. 
\tag{26}
\end{equation}%
Therefore we obtain

\begin{equation}
\eta ^{AB}\frac{\partial x^{\hat{\nu}}}{\partial \xi ^{A}}\frac{\partial x^{%
\hat{\sigma}}}{\partial \xi ^{B}}\gamma _{\hat{\nu}\hat{\sigma}}=\frac{%
\partial x^{\hat{\nu}}}{\partial \lambda ^{pq}}\frac{\partial x^{\hat{\sigma}%
}}{\partial \lambda ^{rs}}\gamma _{\hat{\nu}\hat{\sigma}}\eta ^{AB}\frac{%
\partial \lambda ^{pq}}{\partial \xi ^{A}}\frac{\partial \lambda ^{rs}}{%
\partial \xi ^{B}},  \tag{27}
\end{equation}%
which by using (25) we get

\begin{equation}
\eta ^{AB}\frac{\partial x^{\hat{\nu}}}{\partial \xi ^{A}}\frac{\partial x^{%
\hat{\sigma}}}{\partial \xi ^{B}}\gamma _{\hat{\nu}\hat{\sigma}%
}=2\varepsilon ^{pr}\varepsilon ^{qs}\frac{\partial x^{\hat{\nu}}}{\partial
\lambda ^{pq}}\frac{\partial x^{\hat{\sigma}}}{\partial \lambda ^{rs}}\gamma
_{\hat{\nu}\hat{\sigma}}.  \tag{28}
\end{equation}%
Thus, the transition $\eta ^{AB}\rightarrow G^{AB}$ shall induce the
transitions $\varepsilon ^{pr}\rightarrow \varphi _{1}g^{pr}$ and $%
\varepsilon ^{qs}\rightarrow \varphi _{2}\gamma ^{qs}$ where $g^{pr}$ and $%
\gamma ^{qs}$ are two different nonsymmetric metrics (see section 2) and $%
\varphi _{1}$ and $\varphi _{2}$ are two conformal factors. Observe that the
sum of the number of degrees of freedom of $\varphi _{1}g^{pr}$ and $\varphi
_{2}\gamma ^{qs}$ is equal to the ten degrees of freedom of the symmetric
metric $G^{AB}.$

Considering these preliminaries one finds that the action (4), or (9) which
corresponds to the $1+1-$matrix-brane system, can be written in the
alternative form%
\begin{equation}
S=\frac{1}{2}\tint d\lambda ^{1+1}\sqrt{-g}\sqrt{-\gamma }\left[
g^{ab}\gamma ^{mn}\frac{\partial x^{\hat{\nu}}}{\partial \lambda ^{am}}\frac{%
\partial x^{\hat{\sigma}}}{\partial \lambda ^{an}}\gamma _{\hat{\nu}\hat{%
\sigma}}-2\right] .  \tag{29}
\end{equation}%
This action can be obtained from the quatl action

\begin{equation}
\mathcal{S}=\tint d\tau \left[ \dot{\lambda}^{am}p_{am}-\frac{1}{2}\sqrt{-g}%
\sqrt{-\gamma }(g^{ab}\gamma ^{mn}p_{am}p_{bn}-\frac{2}{S+T})\right] . 
\tag{30}
\end{equation}%
We recognize that (30) has exactly the same form than (10). This suggested
that the coordinates $\lambda ^{am}$ and $\zeta ^{am}$ which describe the
motion of the quatl must be related. In fact, using (2) we find

\begin{equation}
\lambda ^{pq}=\left( 
\begin{array}{cc}
\zeta ^{11}+\zeta ^{12} & \zeta ^{21}+\zeta ^{22} \\ 
\zeta ^{21}-\zeta ^{22} & \zeta ^{11}-\zeta ^{12}%
\end{array}%
\right) ,  \tag{31}
\end{equation}%
which leads to a transformation

\begin{equation}
\lambda ^{pq}=\Sigma _{ab}^{pq}\zeta ^{ab},  \tag{32}
\end{equation}%
where the only nonvanishing components of $\Sigma _{ab}^{pq}$ are 
\begin{equation}
\begin{array}{ccccccc}
\Sigma _{11}^{11}=1, &  & \Sigma _{12}^{11}=1, &  & \Sigma _{21}^{12}=1, & 
& \Sigma _{22}^{12}=1, \\ 
&  &  &  &  &  &  \\ 
\Sigma _{21}^{21}=1, &  & \Sigma _{22}^{21}=-1, &  & \Sigma _{11}^{22}=1, & 
& \Sigma _{12}^{22}=-1.%
\end{array}
\tag{33}
\end{equation}

Substituting (32) into (24) and using the line element (3) we obtain the
formula

\begin{equation}
\frac{1}{2}\Sigma _{ab}^{pq}\Sigma _{cd}^{rs}\varepsilon _{pr}\varepsilon
_{qs}=\eta _{ac}\eta _{bd},  \tag{34}
\end{equation}%
which can be verified using (33).

\bigskip\ 

\noindent \textbf{5.- Final comments}

\smallskip\ 

In this paper we have established a number of connections between the
results of references [15] and [20]. In particular, we have shown that the
Duff's prescription leads to an alternative but equivalent approach to the
quatl theory. It comes to be evident that these results reinforces the idea
proposed in Ref. [15] that the $2+2$ and $1+1$ signatures are exceptional
structures (see Ref. [22] for a motivation on exceptional structures in
mathematics).

From the present work it emerges one more interesting possibility of writing
the Nambu-Goto action in a $2+2$-target spacetime. Let us introduce the
coordinates $y^{ab}$ defined in terms of the coordinates $x^{A}$ as

\begin{equation}
\begin{array}{ccccccc}
y^{11}\equiv x^{1}, &  & y^{22}\equiv x^{2}, &  & y^{12}\equiv x^{3}, &  & 
y^{21}\equiv x^{4}.%
\end{array}
\tag{35}
\end{equation}%
We find the matrix (12) can also be written as

\begin{equation}
h_{ab}=\frac{\partial y^{am}}{\partial \xi ^{a}}\frac{\partial y^{bn}}{%
\partial \xi ^{b}}\eta _{ab}\eta _{mn},  \tag{36}
\end{equation}%
and therefore the Nambu-Goto action can be written as

\begin{equation}
S=\tint d\xi ^{1+1}\sqrt{-\det (h_{ab})},  \tag{37}
\end{equation}%
where

\begin{equation}
\det (h_{ab})\equiv \frac{1}{2!}\varepsilon ^{ic}\varepsilon ^{jd}\eta
^{ef}\eta ^{gh}\eta ^{rs}\eta ^{lm}u_{egi}u_{fhj}u_{rlc}u_{smd},  \tag{38}
\end{equation}%
with

\begin{equation}
u_{pqa}=\frac{\partial y_{pq}}{\partial \xi ^{a}}.  \tag{39}
\end{equation}%
Of course, using the prescription (34) one can prove that the action (37)
and (14) are equivalents. It remains to see whether the action (37) may be
helpful for having a better understanding of the hidden symmetries of the
Nambu-Goto action described by Duff [20].

Let briefly comment on the scenario in which the present discussion may have
physical implications. It turns out that the $2+10$-dimensional spacetime
signature has emerged as one of the most interesting possibilities for the
understanding of both supergravity and super Yang-Mills theory in $D=11$.
For our considerations what is important is that the $2+10$-dimensional
theory seems to be the natural background for the $2+2$-brane (see Ref. [23]
and references therein). Thus, let us think in the possible transition

\begin{equation}
M^{2+10}\rightarrow M^{2+2}\times M^{0+8},  \tag{40}
\end{equation}%
which, in principle, can be achieved by some kind of symmetry breaking
applied to the full metric of the spacetime manifold $M^{2+10}.$ It has been
shown that the symmetry $SL(2,R)$ makes the $2+2$-signature an exceptional
one [15]. On the other hand, the signature $0+8$ is euclidean and in
principle can be treated with the traditional methods such as the octonion
algebraic approach. In pass, it is interesting to observe that octonion
algebra is also exceptional in the sense of the celebrated Hurwitz theorem.
Thus, we see that both $2+2$ and $0+8$ are exceptional signatures. This
means that the transition (40) is physically interesting.

Consider now the action (29) in the particular case of $2+10$-dimensional
target spacetime background

\begin{equation}
S=\frac{1}{2}\tint d\lambda ^{1+1}\sqrt{-g}\sqrt{-\gamma }\left[
g^{ab}\gamma ^{mn}\frac{\partial x^{\hat{\nu}}}{\partial \lambda ^{am}}\frac{%
\partial x^{\hat{\sigma}}}{\partial \lambda ^{an}}\eta _{\hat{\nu}\hat{\sigma%
}}-2\right] ,  \tag{41}
\end{equation}%
where we have replaced the curved metric $\gamma _{\hat{\nu}\hat{\sigma}}$
by the flat metric $\eta _{\hat{\nu}\hat{\sigma}}$ and we assume that the
indices $\hat{\nu},\hat{\sigma}$ now run from $1$ to $12.$Splitting the flat
metric $\eta _{\hat{\nu}\hat{\sigma}}$ according to the transition $%
2+10\rightarrow $ $(2+2)+(0+8)$ we find that (41) can be written as

\begin{equation}
S=S_{1}+S_{2},  \tag{42}
\end{equation}%
where%
\begin{equation}
S_{1}=\frac{1}{2}\tint d\lambda ^{1+1}\sqrt{-g}\sqrt{-\gamma }\left[
g^{ab}\gamma ^{mn}\frac{\partial x^{A}}{\partial \lambda ^{am}}\frac{%
\partial x^{B}}{\partial \lambda ^{bn}}\eta _{AB}\right]  \tag{43}
\end{equation}%
and

\begin{equation}
S_{2}=\frac{1}{2}\tint d\lambda ^{1+1}\sqrt{-g}\sqrt{-\gamma }\left[
g^{ab}\gamma ^{mn}\frac{\partial x^{\hat{A}}}{\partial \lambda ^{am}}\frac{%
\partial x^{\hat{B}}}{\partial \lambda ^{bn}}\eta _{\hat{A}\hat{B}}-2\right]
.  \tag{44}
\end{equation}%
Here, the indices $A,B$ run from $1$ to $4$ and $\hat{A},\hat{B}$ run from $%
5 $ to $12$. Employing (13), we can make the change $x^{A}\rightarrow x^{pq}$
which allows us to write the action (43) in the form

\begin{equation}
S_{1}=\frac{1}{2}\tint d\lambda ^{1+1}\sqrt{-g}\sqrt{-\gamma }\left[
g^{ab}\gamma ^{mn}a_{pqam}a_{rsbn}\varepsilon ^{pr}\varepsilon ^{qs}\right] .
\tag{45}
\end{equation}%
It turns out that the variables $a_{pqam}$ generalize the equivalent Duff's
definition given in the expression (2.4) of Ref. [20]. Thus, one should
expect additional discrete symmetries. In fact, in principle, we should have
eight possible transformations

\begin{equation}
\begin{array}{ccc}
a_{pqam}\rightarrow a_{apqm,} &  & a_{pqam}\rightarrow a_{qapm}, \\ 
&  &  \\ 
a_{pqam}\rightarrow a_{aqpm}, &  & a_{pqam}\rightarrow a_{paqm,} \\ 
&  &  \\ 
a_{pqam}\rightarrow a_{mpqa}, &  & a_{pqam}\rightarrow a_{qmpa}, \\ 
&  &  \\ 
a_{pqam}\rightarrow a_{mqpa}, &  & a_{pqam}\rightarrow a_{pmqa}.%
\end{array}
\tag{46}
\end{equation}%
which represent discrete symmetries. Therefore, we see that at least in the $%
2+2$-dimensional sector of the full $2+10$-dimensional theory the present
development provides a physical application in terms of the extended
discrete symmetries (46).

It may be interesting for further research to understand the present
development from the point of view of supersymmetry and superconformal
group. In particular it appears interesting to investigate $1+1$%
-matrix-brane system from the point of view of the $2$-brane case at the end
of the De Sitter universe due to Batrachenko, Duff and Lu [24]. Finally,
since self-duality seems to be deeply connected with the $2+2$-signature
[1]-[2] one may be interested to find the self-dual aspects of the $1+1$%
-matrix-brane system.

\bigskip\ 

\noindent \textbf{Acknowledgments: }This work was partially supported by
grants PIFI 3.2.

\smallskip\

\end{document}